\pdfoutput=1

\documentclass[aps,pra,amsmath,amssymb,superscriptaddress,twocolumn,nofootinbib]{revtex4-1}

\usepackage{verbatim,bbm,graphicx,color,stmaryrd}
\usepackage[pdfborder={0 0 0}]{hyperref}

\newcommand{\mathd}{\mathrm{d}}

\newcommand{\tmmathbf}[1]{\ensuremath{\bs{#1}}}
\newcommand{\tmop}[1]{\ensuremath{\operatorname{#1}}}

\newcommand{\ie}{{\it i.e.\ }}
\newcommand{\eg}{{\it e.g.\ }}
\newcommand{\msf}{\ensuremath{\mathsf}}
\newcommand{\bra}[1]{\ensuremath{\langle #1 |}}
\newcommand{\ket}[1]{\ensuremath{| #1 \rangle}}
\newcommand{\braket}[2]{\ensuremath{\langle #1| #2 \rangle}}
\newcommand{\bs}{\ensuremath \boldsymbol}

\begin{document}

\title{Optimally Convergent Quantum Jump Expansion}
\author{Felix Lucas}
\affiliation{Max Planck Institute for the Physics of Complex Systems, N\" othnitzer Stra\ss{}e 38, 01187 Dresden, Germany}
\affiliation{University of Duisburg-Essen, Faculty of Physics, Lotharstra\ss{}e 1-21, 47057 Duisburg, Germany}
\author{Klaus Hornberger}
\affiliation{University of Duisburg-Essen, Faculty of Physics, Lotharstra\ss{}e 1-21, 47057 Duisburg, Germany}

\date{\today}

\begin{abstract}
A method for deriving accurate analytic approximations for Markovian open quantum systems was recently introduced in [F.\ Lucas and K.\ Hornberger, Phys.\ Rev.\ Lett.\ \textbf{110}, 240401 (2013)]. Here, we present a detailed derivation of the underlying non-perturbative jump expansion, which involves an adaptive resummation to ensure optimal convergence. Applying this to a set of exemplary master equations, we find that the resummation typically leads to convergence within the lowest two to five orders. Besides facilitating analytic approximations, the optimal jump expansion thus provides a numerical scheme for the efficient simulation of open quantum systems.
\end{abstract}

\maketitle

\section{Introduction}

Virtually all real quantum systems are in contact with their surroundings. The ensuing incoherent dynamics is the key ingredient for explaining such important phenomena as energy dissipation \cite{Weiss1999, Breuer2002}, quantum Brownian motion \cite{Grabert1988, Hornberger2009b}, or quantum measurements \cite{Haroche2007, Wiseman2010}. Moreover, it is the basis of decoherence theory \cite{Zurek2003, Joos2003}, which describes the emergence of classical properties on the macroscopic scale and has so far withstood all experimental tests \cite{Haroche1996,  Hornberger2012}. Recently, incoherent dynamics has also been used as a new lever for controlling open quantum systems \cite{Prezhdo2000, Parkins2003, Zoller2008, Viola2009, Schirmer2010, Polzik2011, Lukin2013}.

Our understanding of these effects is limited by the fact that only a handful of the simplest Markovian open problems can be solved analytically, \eg a two-level system with dephasing or a harmonic oscillator in a zero-temperature bosonic bath \cite{Breuer2002}. Therefore, one usually has to content oneself with their numerical simulation. In spite of the undeniable power of state-of-the-art numerical algorithms \cite{Kosloff2003, Vacchini2007, Eisfeld2009}, such treatments can yield only explicit solutions to model systems with specific parameters.

Recently, we showed that highly accurate analytic approximations for Markovian quantum dynamics can be obtained based on a non-perturbative expansion into periods of continuous evolution and discontinuous jumps \cite{Lucas2013}. Crucially, the convergence of this \emph{jump expansion} can be enhanced via an adaptive resummation. While being simple enough to be treated analytically, the resulting lowest order terms may describe the open system evolution to per mil accuracy \cite{Lucas2013}.

In the present article we provide a detailed derivation of the Dyson-like jump expansion. Focusing on the convergence properties of the series under invariant transformations of the associated master equation,  we derive its optimally convergent resummation. We also show that simpler suboptimal resummations offer considerable freedom to adapt the jump expansion for specific purposes, \emph{e.g.}\ for analytical tractability. The numerical implementation of a set of frequently studied master equations demonstrates that the optimally convergent resummation reliably generates expansions that converge within the lowest two to five orders. This underscores that the expansion has the potential to yield analytic approximations for the dynamics of many different master equations. At the same time, it opens up a new way for the efficient numerical solution of open quantum dynamics, which is in some sense complementary to the quantum trajectory approach. Being applicable to any Markovian system, the resummation method could thus become a versatile addition to the toolbox of open quantum systems.

In Sec.~\ref{sec:jumpexpansion} we first demonstrate how the jump expansion is derived, based on an arbitrary splitting of the Markovian generator. We then parametrize different possible jump expansions and quantify their convergence by means of suitably defined weights of the constituent terms. On this basis, the optimally convergent resummation of the jump expansion is then derived in Sec.~\ref{sec:resummation}. Since its intricate, adaptive structure is not always favorable for practical applications we also outline a general way to derive simpler suboptimal resummations. After a brief description in Sec.~\ref{sec:numericalimp} of how to numerically implement and assess general Markovian dynamics with the jump expansion, we benchmark the convergence with and without the optimal resummation in Sec.~\ref{sec:casestudies}. This is done in three paradigmatic Markovian open systems: the damped harmonic oscillator, the decoherence of a quantum Brownian particle, and a nonselective measurement with feedback. Finally, we present our conclusions in Sec.~\ref{sec:conclusion}.

\section{The Jump Expansion}
\label{sec:jumpexpansion}

\subsection{The Markovian Master Equation}

An open quantum system exhibits Markovian behavior if it is coupled to a memoryless environment, \ie if the environment correlation time is much smaller than the system time scale. In this case, the system dynamics is generated by a time local superoperator $\mathcal L(t)$, as described by the master equation $\partial_t \rho_t = \mathcal L(t) \rho_t$. Solutions of the master equation are expressed in terms of the dynamical propagators $\mathcal U(t,0)$ which map system states from the initial time $t_0=0$ to time $t$, \ie
\begin{equation}
\mathcal U (t,0) \rho_{0}= \mathcal T \exp\left[\int_0^t \mathcal L (t') \mathd t'\right] \rho_0 =\rho_t,
\label{propagator}
\end{equation}
where $\mathcal T$ denotes time-ordering. In order for this time evolution to be physical, \ie completely positive and trace-preserving, the generator $\mathcal L(t)$ must be of the form \cite{Breuer2002}
\begin{equation}
  \mathcal{L}(t) \rho_t =- \frac i \hbar \left[ \msf H, \rho_t \right] + \sum_j \msf L_j \rho_t \msf L_j^{\dag} - \frac{1}{2} \msf L_j^{\dag} \msf L_j \rho_t - \frac{1}{2} \rho_t \msf L_j^{\dag} \msf L_j . \label{meq}
\end{equation}
The commutator with the Hermitian operator $\msf H$ induces unitary dynamics just like in closed quantum systems, whereas the second part, involving the so-called jump operators $\msf L_j$, is encountered only in open systems and induces non-unitary dynamics. Please note that, for brevity, a possible time dependence of the master equation is made explicit only in the superoperator $\mathcal L(t)$ and not in the operators $\msf H$ and $\msf L_j$---a convention followed throughout the rest of the article.

\subsection{Expansion by Decomposing the Generator}

Convergent expansions are an invaluable tool for finding good approximations to solutions of differential equations, and they often help to advance our physical intuition. Here, we show how to find a formal expansion of the time evolution $\rho_t$ under the master equation $\partial_t \rho_t = \mathcal L(t) \rho_t$ into periods of continuous evolution and discontinuous jumps \cite{Holevo2001, Vacchini2005, Hornberger2009}.

To derive the expansion, let us first decompose the generator $\mathcal{L}(t)$ into a sum of any two parts
\begin{equation}
  \mathcal{L}(t) = \mathcal{L}_0(t)  +\mathcal{J}(t). \label{genzerleg}
\end{equation}
At this point, the decomposition is arbitrary---it will be specified as we proceed. A formal solution of the master equation in terms of this decomposition is given by
\begin{equation}
 \rho_t  = \mathcal U_0(t,0) \rho_0 + \int_0^t \mathd t' \,\mathcal U_0 (t, t') \mathcal J(t')  \rho_{t'},
 \label{meqipi}
\end{equation}
as can be verified by taking the time derivative and inserting \eqref{meqipi}. Here, $\mathcal U_0(t,0)$ is the exponential of $\mathcal L_0(t)$, integrated over time, \ie $\mathcal U_0(t,0) = \mathcal T \exp[\int_0^t \mathcal L_0(t') \mathd t']$, in analogy to $\mathcal U(t,0)$ from\ Eq.~\eqref{propagator}. Solving the integral equation \eqref{meqipi} iteratively one obtains the formal expansion $\rho_t = \sum_{n = 0}^\infty \rho_t^{(n)}$ with
\begin{equation}
 \rho_t^{(n)} = \int_0^t \mathd t_n \,\mathcal U_0 (t,t_n) \mathcal J(t_n) \rho_{t_n}^{(n-1)},
 \label{recursjumpterm}
\end{equation}
and $\rho_t^{(0)} = \mathcal U_0(t,0) \rho_0$.  Writing out the terms explicitly, the expansion starts as
\begin{multline}
  \rho_t  =  \left[\mathcal U_0(t,0) + \int_0^t \mathd t_1 \mathcal U_0(t,t_1) \mathcal J(t_1) \mathcal U_0(t_1,0) +\right. \\
  \left. \int_0^t\!\!\! \mathd t_2\!\!\int_0^{t_2}\!\!\!\!\! \mathd t_1 \mathcal U_0(t,t_2) \mathcal J(t_2) \mathcal U_0(t_2,t_1) \mathcal J(t_1)\mathcal U_0(t_1,0) \ldots\right] \rho_0.
\label{jumpexr}
\end{multline}
It decomposes the time evolution $\rho_t$ into a series of density matrix transformations $\mathcal U_0$ and $\mathcal J$, in alternating order. While $\mathcal U_0(t_{i+1},t_i)$ describes continuous propagation within the time intervals $[t_i,t_{i+1}]$, the interspersed superoperators $\mathcal J$ represent abrupt transformations of $\rho_t$ effecting discontinuous jumps at times $t_i$. The jump times $t_i$, in turn, are distributed over the entire propagation interval $[0,t]$ as indicated by multi-integrals over $t_i$. The expansion terms $\rho_t^{(n)}$, which we call \emph{jump terms} in the following, are labelled according to the number of jump super operators $\mathcal J$ involved. All in all, the expansion \eqref{jumpexr} is called the \emph{jump expansion}.

The jump expansion has the characteristic form of a Dyson series if one views $\mathcal U_0$ as the unperturbed evolution and $\mathcal J$ as a perturbation. Its main difference with respect to the Dyson series is the absence of a small parameter. Its convergence properties are therefore doubtful, such that the expansion \eqref{jumpexr} alone is of little practical use. However, we will see that convergence can be ensured even in the absence of a small parameter by choosing  optimal, adaptive decompositions \eqref{genzerleg} of the generator.

\subsubsection*{Freedom in the $\mathcal L$-Decomposition}

So far, the decomposition \eqref{genzerleg} of the generator $\mathcal L$ was completely arbitrary. For practical purposes, in particular for studying the convergence of \eqref{jumpexr} and finding good approximations to $\rho_t$, it is however beneficial to consider those decompositions for which all jump terms $\rho_t^{(n)}$ are (unnormalized) physical density matrices.

In oder to characterize these decompositions, note first that the master equation (\ref{meq}) is invariant under the following transformation of the jump operators and of the Hamiltonian
\begin{align}
 \msf L_j & \rightarrow \msf L_{j, \tmmathbf{\alpha}} = \msf L_j + \alpha_j,   \label{ltrafo}\\
 \msf H & \rightarrow \msf H_{\tmmathbf{\alpha}} = \msf  H - \frac{ i \hbar}{2} \sum_j \left( \alpha_j^{\ast} \msf L_j -
  \alpha_j \msf L_j^{\dag} \right),  \label{htrafo}
\end{align}
with $\tmmathbf{\alpha}= \left( \alpha_1, \ldots, \alpha_N \right) \in
\mathbbm{C}^N$. Second, the $\rho^{(n)}_t$ are physical density matrices if both the continuous propagators and the jump transformations in the jump expansion \eqref{jumpexr} are completely positive maps. This requirement is fulfilled by using the generator decomposition $\mathcal L(t) = \mathcal L_{\bs \alpha}(t) + \mathcal J_{\bs \alpha}(t)$ with
\begin{align}
  \mathcal{J}_{\tmmathbf{\alpha}}(t) \rho & = \sum_j \msf L_{j, \tmmathbf{\alpha}}
  \rho \msf L_{j, \tmmathbf{\alpha}}^{\dag} \equiv \sum_j \mathcal{J}_{j,
  \tmmathbf{\alpha}}(t) \rho  \label{jump0},\\
  \mathcal{L}_{\tmmathbf{\alpha}}(t) \rho & = - \frac i \hbar \left\llbracket
 \msf H_{\tmmathbf{\alpha}}^{\text{eff}}(t), \rho \right\rrbracket,
  \label{det0}
\end{align}
and the resulting continuous propagator $\mathcal U_{\bs \alpha}(t_{i+1},t_i) = \mathcal T \exp [ \int_{t_i}^{t_{i+1}} \mathcal L_{\bs \alpha} (t') \mathd t']$. Here, we have introduced the non-hermitian effective Hamiltonian $\msf H_{\tmmathbf{\alpha}}^{\text{eff}}(t) = \msf H_{\tmmathbf{\alpha}} - \frac{i \hbar}{2} \sum_j \msf L_{j, \tmmathbf{\alpha}}^{\dag} \msf L_{j, \tmmathbf{\alpha}}$ and the non-standard commutator $\llbracket \msf A, \msf B\rrbracket = \msf A \msf B - \msf B^\dag \msf A^\dag$. Note that in Eq.~\eqref{jump0} we also decompose the jump superoperator $\mathcal J_{\bs \alpha}$ into different \emph{types} of jumps $\mathcal{J}_{j,\tmmathbf{\alpha}}$ labelled by the jump index $j$. A jump of type $j$ is simply performed by applying the jump operator $\msf L_j$ from the left and $\msf L_j^\dag$ from the right. For this $\mathcal L$-decomposition, the jump terms $\rho_t^{(n)}$ read
\begin{align}
\rho_t^{(n)}  &= \sum_{j_n} \int_0^t \mathd t_n \,\mathcal U_{\bs \alpha} (t,t_n) \mathcal J_{j_n,\bs \alpha} (t_n) \rho_{t_n}^{(n-1)} \nonumber \\
& = \sum_{j_n} \int_0^t \mathd t_n \mathcal T \exp\left[- \frac i\hbar \int_{t_n}^t \msf H^\text{eff}_{\bs \alpha} (t')\mathd t'\right]\mathsf L_{j_n,\bs \alpha} \nonumber \\
& \phantom{=} \times \rho_{t_n}^{(n-1)} \msf L_{j_n,\bs \alpha}^\dag\mathcal T \exp\left[ \frac i\hbar \int_{t_n}^t \msf H^{\text{eff}\, \dag}_{\bs \alpha} (t')\mathd t'\right],
\label{jumpexnat}
\end{align}
and $\rho_t^{(0)} = \mathcal U_{\bs \alpha}(t,0) \rho_0$. The explicit expression for $\rho_t^{(n)}$ involves the multi-integral over $n$ jump times $t_1,\ldots,t_n$, as in Eq.~\eqref{jumpexr}, and in addition a sum over $n$ jump indeces $j_1,\ldots,j_n$.

Inspecting Eqs.~\eqref{ltrafo}--\eqref{det0}, we see that tuples of complex numbers $\bs \alpha$ parametrize those $\mathcal L$-decompositions that lead to physical jump terms $\rho_t^{(n)}$. The parametrizations are equivalent in the sense that any choice of $\bs \alpha$ leads to the same solution of the master equation~\eqref{meq}. This freedom to decompose the generator in different, physically equivalent ways will allow us in the following to enhance the convergence of the jump expansion by optimizing $\bs \alpha$.

\section{Adaptive Resummation of the Jump Expansion}
\label{sec:resummation}

\subsection{Convergence of the Jump Expansion}

One way to characterize the convergence of an expansion is to consider the weights of its constituent terms. In the present case of the jump expansion, it is natural to define the weight of the $n$-th term as
\begin{equation}
  w_n \left( t \right) \equiv \tmop{Tr} \rho^{\left( n \right)}_t  \label{weight},
\end{equation}
such that the weights add up to $\sum _n w_n(t) =\tmop{Tr} \rho_t = 1$. The $w_n (t)$ are positive real numbers since the jump terms $\rho^{\left( n\right)}_t$ in (\ref{jumpexnat}) correspond to positive density matrices. The convergence of \eqref{jumpexr} is \emph{optimal} if the weights of the lowest order terms are maximal at all times. In the following, this is achieved by optimizing over the implicit $\bs \alpha$-dependence of $w_n(t)$.

This optimization problem can be solved by two considerations. First, the fact that $\exp(\mathcal L_{\bs {\alpha}} t)$ is the identity for $t=0$ yields the initial condition $w_n \left( 0 \right) = \delta_{n, 0}$. Second, the time evolution of the weights cascades in the sense that the $n$-th order term gains weight from the $(n-1)$-th order, while loosing weight to the $(n+1)$-th order. As we will see shortly, this cascading time evolution can be described in terms of a suitably defined positive rate operator $\msf \Gamma_{\bs \alpha}$ as
\begin{equation}
 \partial_t w_n(t) = \partial_t \tmop{Tr} \left[ \rho^{(n)}_t\right] = \tmop{Tr}\left[\msf \Gamma_{\bs \alpha} \rho_t^{(n-1)}\right] - \tmop{Tr}\left[\msf \Gamma_{\bs \alpha} \rho_t^{(n)}\right].
 \label{cascade}
\end{equation}
The expectation values $\tmop{Tr}[\msf \Gamma_{\bs \alpha}\rho_t^{(n)}]$ are thus positive real numbers describing the absolute rates at which weight is transferred from $\rho_t^{(n)}$ to $\rho_t^{(n+1)}$. Since the latter involve $n$ and $n+1$ jump super operators, respectively (\emph{cf.}\ Eq.~\eqref{jumpexr}), we call the $\tmop{Tr}[\msf \Gamma_{\bs \alpha}\rho_t^{(n)}]$ \emph{jump rates}. The above considerations together imply that the lowest order weights are maximized by minimizing the jump rates at all times.
 
 To see that the weights $w_n(t)$ evolve as described by Eq.~\eqref{cascade}, we take the time derivative of Eq.~\eqref{recursjumpterm} with a general parametrization $\bs \alpha$,
 \begin{equation}
 \begin{split}
  \partial_t \rho_t^{(n)} \!&= \mathcal U_{\bs \alpha} (t,t) \mathcal J_{\bs{\alpha}}(t) \rho_{t}^{(n-1)}\! +\! \int_0^t\!\!\! \partial_t \mathcal U_{\bs \alpha} (t,t_n) \mathcal J_{\bs{\alpha}}(t_n) \rho_{t_n}^{(n-1)}  \mathd t_n \\
  & = \mathcal J_{\bs{\alpha}}(t) \rho_t^{(n-1)} + \mathcal L_{\bs{\alpha}}(t) \rho_t^{(n)},
  \label{rhonderi}
  \end{split}
 \end{equation}
where the first term stems from the derivative of the integration limit.
Using Eq.~\eqref{rhonderi} in the time derivative of $w_n(t)$, we can insert the definitions \eqref{jump0}, \eqref{det0}, to obtain
\begin{equation}
\begin{split}
  \partial_t w_n(t) &= \tmop{Tr}\left[\sum_j \msf L_{j,\bs{\alpha}} \rho_t^{(n-1)} \msf L_{j,\bs{\alpha}}^\dag-\frac i \hbar \left\llbracket\msf H^\text{eff}_{\bs{\alpha}}(t),\rho_t^{(n)}\right\rrbracket\right] \\
  & = \tmop{Tr}\left[\sum_j \msf L_{j,\bs{\alpha}}^\dag \msf L_{j,\bs{\alpha}} \rho_t^{(n-1)} - \sum_j \msf L_{j,\bs{\alpha}}^\dag \msf L_{j,\bs{\alpha}} \rho_t^{(n)}\right],
  \label{dwn}
 \end{split}
 \end{equation}
 where we have used the cyclic invariance of the trace. Comparing Eqs. \eqref{cascade} and \eqref{dwn}, we see that the rate operator $\msf \Gamma_{\bs \alpha}$ reads
\begin{equation}
\msf \Gamma_{\bs \alpha} = \sum_j \msf L_{j,\bs{\alpha}}^\dagger \msf L_{j,\bs {\alpha}} \equiv \sum_j \msf \Gamma_{j,\bs \alpha}
\label{rateop},
\end{equation}
with the additional splitting of $\msf \Gamma_{\bs \alpha}$ into the rate operators $\msf \Gamma_{j, \bs \alpha}$ associated with different jump types.

\subsection{The Jump Record}

Before we proceed to minimize the jump rates $\tmop{Tr}[\msf \Gamma_{\bs \alpha}\rho_t^{(n)}]$, we must account for the fact that each $\rho_t^{(n)}$ given by Eq.~\eqref{jumpexnat} can be further decomposed into a sum over the jump indices $j_1, \ldots ,j_n$ and a multi-integral over the jump times $t_1, \ldots ,t_n$, \ie
\begin{equation}
 \rho_t^{(n)} = \sum_{j_1 \ldots j_n} \int_0^t \mathd t_n \ldots \int_0^{t_2} \mathd t_1\rho_t^{(\mathfrak R^n)} \equiv \sum_{\{\mathfrak R^n\} }\rho_t^{(\mathfrak R^n)},
 \label{jtermdeco}
\end{equation}
with
\begin{equation}
\rho_t^{(\mathfrak R^n)} \equiv  \mathcal U_{\bs \alpha} (t,t_n) \mathcal J_{j_n,\bs \alpha}(t_n) \ldots \mathcal J_{j_1,\bs \alpha}(t_1) \mathcal U_{\bs \alpha} (t_1,0) \rho_0.
\label{branches}
\end{equation}
Here, the symbol $\sum_{\{\mathfrak R^n\} }$ serves as an abbreviation of the multi-integral and sum. Each constituent $\rho_t^{(\mathfrak R^n)}$ is unambiguously labelled by the \emph{jump record}
\begin{equation}
  \mathfrak{R}^n \equiv (j_1, t_1 ; j_2, t_2 \ldots ; j_n t_n).
\end{equation}
It collects the first $n$ jump labels and jump times and therefore the necessary information to reconstruct one particular realization \eqref{branches} of the state transformations (continuous evolutions and jumps) which make up the jump expansion. Due to the complete positivity of $\mathcal U_{\bs \alpha}$ and $\mathcal J_{j,\bs \alpha}$, the constituents $\rho_t^{(\mathfrak R^n)}$ of $\rho_t^{(n)}$ are physical density matrices in their own right. We call them record-conditioned \emph{branches} (see the schematic diagram, Fig.~\ref{fig:jumptree}).

Given the decomposition of the jump terms $\rho_t^{(n)}$ into branches $\rho_t^{(\mathfrak R^n)}$, the jump rate $\tmop{Tr}[\msf \Gamma_{\bs \alpha}\rho_t^{(n)}]$ can be decomposed into a sum over the record-specific partial jump rates $\tmop{Tr}[\msf \Gamma_{j,\bs \alpha} \rho_t^{(\mathfrak R^n)}]$, \ie
\begin{align}
\tmop{Tr}\left[\msf \Gamma_{\bs \alpha} \rho_t^{(n)} \right] &= \tmop{Tr}\left[\sum_j\msf \Gamma_{j,\bs \alpha} \sum_{\{\mathfrak R^n\} } \rho_t^{(\mathfrak R^n)} \right] \nonumber \\
&\equiv \sum_{j,\{\mathfrak R^n\} } \tmop{Tr}\left[\msf \Gamma_{j,\bs \alpha} \rho_t^{(\mathfrak R^n)} \right].
\end{align}
In order to minimize $\tmop{Tr}[\msf \Gamma_{\bs \alpha}\rho_t^{(n)}]$, all individual $\tmop{Tr}[\msf \Gamma_{j,\bs \alpha} \rho_t^{(\mathfrak R^n)}]$ must be minimized separately. Insertion of the definition \eqref{ltrafo} for $\msf L_{j,\bs{\alpha}}$ yields for the partial jump rates
\begin{align}
 \tmop{Tr}[\msf \Gamma_{j, \bs \alpha} \rho_t^{(\mathfrak R^n)}] &= \tmop{Tr}\left[ (\msf L_j^\dagger \msf L_j  + |\alpha_j|^2 ) \rho_t^{(\mathfrak R^n)}\right] \nonumber\\
  & \phantom{\mathrel{=}} + 2\tmop{Re}\left(\alpha_j^\ast \tmop{Tr}\left[ \mathsf L_j \rho_t^{(\mathfrak R^n)}\right]\right),
 \label{partialrates}
\end{align}
where the first term is always positive and the second term can be positive or negative depending on $\bs \alpha$.

\subsection{The Optimal Adaptive Resummation}
\label{sec:optimalresummation}

\begin{figure}[t]
   \centering
   \includegraphics[width=1\columnwidth]{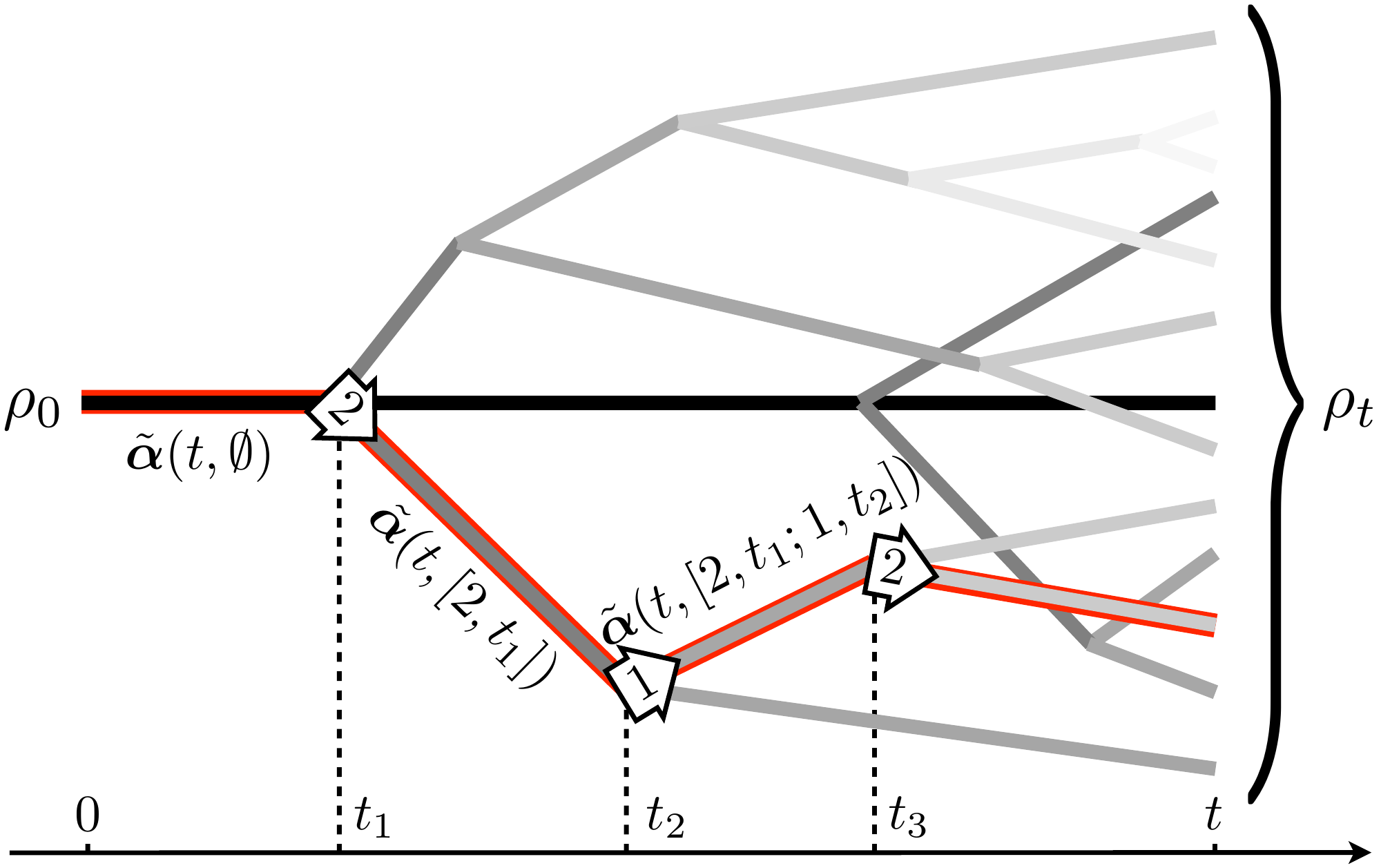}
   \caption{(Color online) Schematic diagram of the jump expansion and the optimal complex shifts $\tilde{\bs \alpha}$: The time evolution from $\rho_0$ to $\rho_t$ is decomposed into branches that consist of a series of continuous and jump-transformations, $\mathcal U_{\bs \alpha}(t_{i+1},t_i)$ and $\mathcal J_{j,\bs \alpha}$, represented by straight lines and branching points, respectively. The branches are distinguished by the times $t_i$ and the types $j$ of the involved jump transformations, \ie by the jump record $\mathfrak R^n$ (see exemplary branch $\mathfrak R^3 = (2,t_1;1,t_2;2,t_3)$ traced in the picture, with $t_i$ and $j$ marked on the time axis on the bottom and inside the arrows at the branch points, respectively). $\rho_t$ is given by the sum over all possible branches $\rho_t^{(\mathfrak R^n)}$, see Eqs.~\eqref{jtermdeco} and \eqref{branches}. The convergence of this expansion is optimized or, in other words, the number of branches that contribute appreciably to $\rho_t$ is minimized by adapting the transformations  $\mathcal U_{\bs \alpha}(t_{i+1},t_i)$ and $\mathcal J_{j,\bs \alpha}$ to each specific branch by means of the time-dependent complex shifts $\bs \alpha = \tilde{\bs \alpha}(t,\mathfrak R^n)$, see Eq.~\eqref{optalpha} and Eqs.~\eqref{jump0} and \eqref{det0}. $\tilde{\bs \alpha}(t,\emptyset)$, $\tilde{\bs \alpha}(t,[2,t_1])$, $\tilde{\bs \alpha}(t,[2,t_1;1,t_2])$, and $\tilde{\bs \alpha}(t,[2,t_1;1,t_2;2,t_3])$ are used in the respective intervals $[t_i,t_{i+1}]$ in case of the marked exemplary branch.}
   \label{fig:jumptree}
\end{figure}

We argued that the jump expansion with optimal convergence is found by minimizing all partial jump rates $\tmop{Tr}[\msf \Gamma_{j, \bs \alpha} \rho_t^{(\mathfrak R^n)}]$ simultaneously at all times. To minimize $\tmop{Tr}[\msf \Gamma_{j, \bs \alpha} \rho_t^{(\mathfrak R^n)}]$, the second term of Eq.~\eqref{partialrates} must be as negative as possible. Denoting by $\tilde{\bs \alpha}$ the optimal choice for $\bs \alpha$, this implies that $\tmop{arg}(\tilde\alpha_j^\ast) = \pi + \tmop{arg}(\tmop{Tr}[\msf  L_j \rho_t^{(\mathfrak R^n)}])$. Using this relation, we can recast Eq.~\eqref{partialrates} at the optimum as a quadratic expression in $|\tilde\alpha_j|$
\begin{align}
 \tmop{Tr}[\msf \Gamma_{j, \tilde{\bs \alpha}} \rho_t^{(\mathfrak R^n)}] &= \tmop{Tr}\left[ (\msf L_j^\dagger \msf L_j  + |\tilde\alpha_j|^2 ) \rho_t^{(\mathfrak R^n)}\right] \nonumber \\
 & \phantom{=} -  2 |\tilde \alpha_j| \left|\tmop{Tr}\left[ \mathsf L_j \rho_t^{(\mathfrak R^n)}\right]\right|.
 \label{tilde}
\end{align}
It is minimal for $|\tilde \alpha_j| = | \tmop{Tr}[\msf L_j \rho_t^{(\mathfrak R^n)}]|/\tmop{Tr}\rho_t^{(\mathfrak R^n)}$. Respecting the above phase requirement, the minimum in \eqref{partialrates} is attained for
\begin{equation}
 \tilde \alpha_j (t,\mathfrak R^n) = - \frac 1 {\tmop{Tr}\rho_t^{(\mathfrak R^n)}}{ \tmop{Tr}\left[\msf L_j \rho_t^{(\mathfrak R^n)}\right]},
 \label{optalpha}
\end{equation}
which depends both on time $t$ and on the jump record $\mathfrak R^n$ of the considered branch. The resulting minimal partial jump rate is given by
\begin{equation}
 \tmop{Tr}[\msf \Gamma_{j, \tilde{\bs \alpha}} \rho_t^{(\mathfrak R^n)}] = \tmop{Tr}\left[\msf L_j^\dagger \msf L_j \rho_t^{(\mathfrak R^n)} \right] - \frac 1 {\tmop{Tr} \rho_t^{(\mathfrak R^n)} }{ \left|\tmop{Tr}\left[\msf L_j \rho_t^{(\mathfrak R^n)} \right]\right|^2}.
 \label{jumprateopt}
\end{equation}

In the previous section we observed that the jump expansion is composed of different realizations of continuous and jump transformations ($\mathcal U_{\bs \alpha}$ and $\mathcal J_{j, \bs \alpha}$) of the initial state, \ie different branches (see Eqs.~\eqref{jtermdeco} and \eqref{branches}, and the schematic diagram, Fig.~\ref{fig:jumptree}). Equation~\eqref{optalpha}, in conjunction with Eqs.~\eqref{jump0} and \eqref{det0}, now shows that its convergence is optimized by adapting the applied transformations to each specific branch. Specifically, the jump operator $\msf L_{j,\tilde{\bs \alpha}}$ that defines the optimal $\mathcal J_{j,\tilde{\bs \alpha}}$ acquires a shift, given by its expectation value in the respective branch. After a jump event of type $j$, the $\mathcal J_{j,\tilde{\bs \alpha}}$ are therefore different from the $\mathcal J_{j',\tilde{\bs \alpha}}$ after a jump event of type $j'$. The optimal shifts $\tilde{\bs \alpha}$ are hence updated at such branching points depending on the type of jump and on the time when it takes place. After a series of $n$ jumps $\tilde{\bs \alpha}$ depends on the complete record $\mathfrak R^n$ of past jumps.

To summarize, while Dyson-like expansions are usually defined in terms of a fixed decomposition of the generator we here consider an expansion in which the decomposition changes from one term to the next. Formally, the transformation of the jump expansion to its optimally convergent form can be considered an adaptive resummation in the above sense that it adapts the $\mathcal L$-decomposition both to time and to each specific branch.

In view of the fact that the jump expansion was derived in analogy to a usual Dyson series in Sec.~\ref{sec:jumpexpansion}, it is not obvious that record conditioned, time dependent decompositions $\mathcal L(t) = \mathcal L_{\tilde{\bs{\alpha}} (t, \mathfrak{R}^n)}(t) + \mathcal J_{\tilde{\bs{\alpha}} (t, \mathfrak{R}^n)}(t)$ generate valid solutions of the master equation. To see that this is indeed the case, we take the time derivative of $\rho_t^{(n)}$ just as in Eq.~\eqref{rhonderi} and use Eqs.~\eqref{jtermdeco} and \eqref{branches} to obtain
\begin{equation}
\begin{split}
 \partial_t \rho_t^{(n)} &= \sum_{j_n, \{ \mathfrak{R}^{n - 1} \}} \mathcal{J}_{j_n,   \tilde{\tmmathbf{\alpha}} (t, \mathfrak{R}^{n - 1})}(t)  \rho_t^{(\mathfrak{R}^{n - 1})} \\
 &\phantom{\mathrel{=}} + \sum_{\{ \mathfrak{R}^n \}}  \mathcal{L}_{\tilde{\tmmathbf{\alpha}} (t, \mathfrak{R}^n)} (t)  \rho_t^{(\mathfrak{R}^n)}.
 \end{split}
\end{equation}
When combining the second summand of $\partial_t \rho^{(n)}_t$ and the first summand of $\partial_t \rho^{(n + 1)} _t$ we obtain the $\bs \alpha$-independent generator as prefactor, \ie $\mathcal{L} (t) \sum_{\{ \mathfrak{R}^n \}} \rho^{(\mathfrak{R}^n)}_t$, so that the sum over the $\mathfrak R^n$-conditioned branches gives $\rho_t^{(n)}$. The sum over all orders then yields the master equation.

\subsubsection*{Conditions for Vanishing Jump Rates}

The jump operators $\msf L_{j,\alpha}$ with the complex shifts $\tilde{\bs{\alpha}} (t, \mathfrak{R}^n)$ given by Eq.~\eqref{optalpha} lead to an optimally convergent jump expansion. Moreover, since the jump rates $\tmop{Tr}[\msf \Gamma_{j, \tilde{\bs \alpha}} \rho_t^{(\mathfrak R^n)}]$ (Eq.~\eqref{jumprateopt}) quantifiy how fast higher order terms get populated over time, we can conclude that the smaller the optimal jump rates, the better the convergence of the optimized jump expansion. For non-vanishing jump rates, the number of terms necessary to approximate $\rho_t$ with a given accuracy increases with time $t$. If, however, all jump rates vanish after some time, the weights of the expansion terms remain constant and it is safe to approximate $\rho_t$ with a fixed number of terms for all $t$. Let us therefore examine the conditions for $\tmop{Tr}[\msf \Gamma_{j, \tilde{\bs \alpha}} \rho_t^{(\mathfrak R^n)}]$ to vanish.

For pure states $\rho_t^{(\mathfrak R^n)} = \ket \varphi \bra \varphi$ and letting $\ket \chi \equiv \msf L_j \ket{\varphi}$, the condition $\tmop{Tr}[\msf \Gamma_{j, \tilde{\bs \alpha}} \rho_t^{(\mathfrak R^n)}] = 0$ in Eq.~\eqref{jumprateopt} leads to $\braket \varphi \chi \braket \chi \varphi = \braket \varphi \varphi \braket \chi \chi$. Clearly, this is fulfilled if  $\ket \chi = c \ket \varphi$, \ie if $\ket{\varphi}$ is an eigenvector of $\msf L_j$. Similarly for mixed states $\rho_t^{(\mathfrak R^n)} = \sum_i p_i \ket{\varphi_i} \bra{\varphi_i}$ in diagonal form, we find that all $\ket{\varphi_i}$ must be eigenvectors of $\msf L_j$ to the \emph{same} eigenvalue.
What is more, this suggests that in order for all $\tmop{Tr}[\msf \Gamma_{j, \tilde{\bs \alpha}} \rho_t^{(\mathfrak R^n)}]$ to vanish \emph{simultaneously}, the $\rho_t^{(\mathfrak R^n)}$ must be composed of simultaneous eigenvectors of all $\msf L_j$, which implies that all $\msf L_j$ must commute. In that case, the jump operators define a preferred basis for decoherence.

One may now argue that the above conditions will never be fulfilled in practice because the simultaneous eigenstates of all $\msf L_j$ form a  set of measure zero within state space. Various studies indicate, however, that in many open quantum problems any initial state quickly evolves towards a mixture of preferred basis states, or pointer states, if the latter exist \cite{Gisin1995, Diosi2000, Hornberger2009c, *Hornberger2010}. The resulting jump rates are then either zero or close to zero, depending on the influence of the Hamiltonian $\msf H$ on the basis states.

\subsection{Suboptimal Resummations}

The biggest obstacle to an analytical or numerical assessment of the optimal adaptive jump expansion is the need to evaluate the jump time multi-integrals, see Eq.~\eqref{jumpexnat}. In particular, the record-dependent complex shifts $\tilde{\bs \alpha}(t,\mathfrak R^n)$ imply that the jump and the continuous evolution superoperators in the integrand depend on all previous jump times. Hence, for the purpose of deriving analytic approximations \cite{Lucas2013} or implementing efficient numerical algorithms for open quantum dynamics, it may be advisable to sacrifice some of the convergence of the optimal resummation for a simpler algebraic structure of the integrand.

A simpler, suboptimal resummation can be obtained from the optimal one by disregarding some of the parameters in the jump record $\mathfrak R^n$ that $\tilde{\bs \alpha}$ depends on. We saw that $\mathfrak R^n$ labels the branches of the jump expansion unambiguously and that the convergence is optimized by minimizing the jump rates in each branch individually. In contrast, removing some parameters in $\mathfrak R^n$ implies that we \emph{group} different branches and optimize the \emph{joint} jump rate of all grouped branches.

For example, one can minimize the jump rates $\tmop{Tr}[\msf \Gamma_{\bs \alpha} \rho_t^{(n)}]$ in Eq.~\eqref{cascade} without resorting to the partial rates $\tmop{Tr}[\msf \Gamma_{j, \bs \alpha} \rho_t^{(\mathfrak R^n)}]$, which is equivalent to grouping all branches with the same number $n$ of jumps. After inserting the definitions \eqref{rateop} and \eqref{ltrafo} for $\msf \Gamma_{\bs \alpha}$ and $\msf L_{\bs \alpha}$ into Eq.~\eqref{cascade}, the optimization is carried out in the same way as in the previous section in Eqs.~\eqref{partialrates}--\eqref{optalpha}. As a result, the suboptimal complex shifts read
\begin{equation}
 \alpha_j(t,n) = - \frac 1 {\tmop{Tr} \rho_t^{(n)}}\tmop{Tr} \left[\msf L_j \rho_t^{(n)}\right].
 \label{nalpha}
\end{equation}

Another suboptimal resummation is realized by updating the jump operators only after a jump has occurred and leaving them constant until the next jump takes place, \ie
\begin{equation}
 \bs \alpha(t,\mathfrak R^n) \equiv \tilde{\bs \alpha} (t_n,\mathfrak R^n), \text{ for } t \in [t_n,t_{n+1}).
 \label{tconstalpha}
\end{equation}
This eliminates the continuous $t$-dependence of $\tilde{\bs \alpha}$.

One may further remove the dependence on the jump times $t_1 \ldots t_n$. In this case, the shifts of the jump operators are only conditioned on the jump \emph{indices} $\mathfrak j^n = (j_1,\ldots,j_n)$ of the previous jumps. This has proved particularly useful for deriving analytic approximations for the process of particle-detection and the open Landau-Zener system \cite{Lucas2013}. Without knowledge of the jump time intervals we cannot specify the continuous evolution operators $\mathcal U_{\bs \alpha}(t_{i+1},t_i)$ involved in the jump expansion. Hence, we must assume complete ignorance of the state before the first jump $\rho_{t_1} \propto \mathbbm{1}$. One can, however, use the jump index $j_1$ to approximate the state after the first jump as $\rho_{t_1}' \propto\msf L_{j_1,  \tmmathbf{\alpha}} \mathbbm 1 \msf L_{j_1,  \tmmathbf{\alpha}}^{\dag}$. We are thus led to update the complex shifts $\bs \alpha$ using $\rho_{t_1}'$ in the same way we used $\rho_t^{(n)}$ in Eq.~\eqref{nalpha}. The complex shift after $n$ jumps is then given by
\begin{equation}
 \alpha_j (\mathfrak j ^n) = - \frac{\tmop{Tr} (\msf L_j \msf L_{j_n,  \tmmathbf{\alpha} (\mathfrak j^{n - 1})} \msf L_{j_n,  \tmmathbf{\alpha} (\mathfrak j^{n - 1})}^{\dag})}{\tmop{Tr} (\msf L_{j_n,  \tmmathbf{\alpha} (\mathfrak j^{n - 1})} \msf L_{j_n,  \tmmathbf{\alpha} (\mathfrak j^{n - 1})}^{\dag})} . \label{bestalpha}
\end{equation}

In order to obtain analytically tractable approximations for specific systems (such as in Ref.~\cite{Lucas2013}), one needs to select the adequate suboptimal resummation carefully. The removal of parameters as outlined above yields simplified jump time integrals, and one must ensure that the resulting convergence properties are still comparable to those of the optimal resummation. The generally very strong convergence of the optimal resummation that we observed, see Sec.~\ref{sec:casestudies}, poses an upper bound for the convergence of any suboptimal resummation. This suggests that one has a considerable margin to find highly convergent, analytically tractable resummations for any Markovian master equation.

\subsection{Relation to Quantum Trajectories}

The jump expansion formalism derived above is in some aspects similar to the unraveling of master equations based on quantum trajectories \cite{Knight1998}. In particular, jump operators $\msf L_j$ that are shifted by their respective trajectory-conditioned expectation values, similar to Eq.~\eqref{optalpha}, were found to generate trajectories with interesting properties \cite{Diosi1986, Wiseman1995, Hornberger2009c, *Hornberger2010}. Let us therefore compare the two approaches.

An unraveling of a master equation \eqref{meq} describes the solutions $\rho_t$ as an ensemble average over \emph{pure state} quantum trajectories $\ket{\psi_t}$, whose time evolution is described by a stochastic Schr\"odinger equation. The jump expansion, in contrast, decomposes $\rho_t$ into a sum over constituent \emph{mixed states} $\rho_t^{(n)}$. They are associated with the terms of a Dyson-like expansion of the deterministic time evolution of $\rho_t$ under the master equation.

If we choose to decompose the evolution into more than just two types of transformations (unperturbed evolution and perturbations in case of the usual Dyson series), we can further break down $\rho_t^{(n)}$, at most up to the record-conditioned branches $\rho_t^{(\mathfrak R^n)}$. For pure initial states $\rho_0 = | \psi_0 \rangle \langle \psi_0 |$, the branches remain pure states, $\rho_t^{(\mathfrak R^n)} = \ket{\psi^{(\mathfrak{R}^n)}_t } \bra{\psi^{(\mathfrak{R}^n)}_t }$, given by
\begin{equation}
  | \psi^{(\mathfrak{R}^n)}_t \rangle = \mathcal T \exp\left[- \frac{i}{\hbar}\int_{t_n}^t \mathsf{H}_{\bs \alpha}^{\text{eff}} (t') \mathd t' \right] \mathsf{L}_{j_n, \tmmathbf{\alpha}}(t_n) \ket{\psi_{t_n}^{(\mathfrak R^{n-1})}}, \label{traject}
\end{equation}
with $\ket{ \psi^{(\mathfrak{R}^0)}_t } = \mathcal T \exp [- i/\hbar\int_0^t \mathsf{H}_{\bs \alpha}^{\text{eff}} (t') \mathd t' ]\ket{\psi_0}$, see Eq.~\eqref{branches}. In this case, and for a \emph{fixed} choice of the complex shifts $\bs \alpha$, each branch $\rho_t^{(\mathfrak R^n)}$ is equivalent to a particular quantum trajectory when normalized with its respective weight $\braket{\psi^{(\mathfrak{R}^n)}_t }{\psi^{(\mathfrak{R}^n)}_t }$. We can then regard the jump expansion as a sum over pure state quantum trajectories, where each trajectory is weighted with the probability of its occurrence \cite{Hornberger2009}.

The distinctive feature of the jump expansion is the ability to lump together different branches. The resulting terms are then necessarily mixed states, even for an initially pure state. This gives us the freedom to apply general resummations (\ie reorder and regroup the branches arbitrarily) in order to obtain a new expansion that is, for example, strongly convergent or analytically accessible. For instance, the suboptimal resummations \eqref{nalpha}, \eqref{tconstalpha}, and \eqref{bestalpha} group branches with the same number of jumps or with the same sequence of jump indices by making use of only partial information of the jump record $\mathfrak R^n$. 

As mentioned above, the conditional update rule \eqref{optalpha} requiring complete knowledge of $\mathfrak R^n$ has been discussed already in other contexts. The corresponding jump operators were used to map the trajectory to orthogonal subspaces \cite{Diosi1986} or to minimize the algorithmic information of a fictitious measurement \cite{Wiseman1995}. They were also employed to identify the pointer basis of a given open quantum system \cite{Hornberger2009c, *Hornberger2010}. In the present case, the purpose of the operators $\msf L_{j,\bs \alpha}$ with \eqref{optalpha} is quite different: They determine an optimally convergent expansion of the open system dynamics $\rho_t$.

\section{Numerical Implementation}
\label{sec:numericalimp}

Having derived the jump expansion and its resummations, let us spell out an efficient method to implement them numerically. It uses a \emph{classical} Monte Carlo integration algorithm for the iterative approximation of the $n$-dimensional integrals $\rho_t^{(n)}$ that contribute to the solutions $\rho_t = \sum_n \rho_t^{(n)}$ of the master equation~\eqref{meq}. Since the resummations attribute large weights to the leading order terms, one can assess $\rho_t$ term by term in the order of their importance. Having calculated $m$ terms one can also estimate how much the next term could contribute at most to $\rho_t$, since the sum of the weights is always unity. The standard numerical quantum trajectory method \cite{Breuer2002,Knight1998}, in contrast, approximates $\rho_t$ by a single indiscriminate sum of \emph{pure} states which are weighted with a static, predetermined probability distribution.

The terms $\rho_t^{(n)}$ are generated from the initial state $\rho_0$ by a series of time-dependent transformations, see  Eq.~\eqref{jumpexnat}. For a given set of jump times $\mathfrak t^n = (t_1,\ldots, t_n)$, they take the form
\begin{equation}
 \rho_t^{(\mathfrak t^n)} \equiv \mathcal U_{\bs \alpha}(t,t_n) \mathcal J_{\bs \alpha}(t_n) \ldots \mathcal J_{\bs \alpha}(t_1)\mathcal U_{\bs \alpha}(t_1,0) \rho_0,
 \label{tdtrafo}
\end{equation}
with $\mathcal U_{\bs \alpha}$ and $\mathcal J_{\bs \alpha}$ completely specified by Eqs.~\eqref{jump0} and \eqref{det0} and the choice of $\bs \alpha$ (we have $\bs \alpha = 0$ for the jump expansion without resummation and $\bs \alpha = \tilde{\bs \alpha}$ for the optimal resummation, see Eq.~\eqref{optalpha}). The $n$-th order term $\rho_t^{(n)}$ is given by the multi-integral over the jump times
\begin{equation}
 \rho_t^{(n)} = \int_0^t \mathd t_n \ldots \int_0^{t_2} \mathd t_1 \rho_t^{(\mathfrak t^n)},
 \label{mcintterm}
\end{equation}
which can be viewed as an average over all possible transformations \eqref{tdtrafo}.

The method of choice for the numerical approximation of multi-integrals is Monte Carlo integration \cite{NumericalRecipes}. It prescribes that the integral of a function $f(\mathfrak t^n)$ over $\mathfrak t^n$ can be approximated by the arithmetic mean of $V f(\mathfrak t_i^n)$, $i = 1,\ldots, N$,
\begin{equation}
  \int_V f(\mathfrak t^n) \mathd \mathfrak t^n \approx \frac V N \sum_{i=1}^N f(\mathfrak t_i^n ),
\end{equation}
where $\mathfrak t_i^n$ represents the $i$-th sample of jump-times $ (t_1, \ldots, t_n)$ drawn from a uniform distribution, and $V$ is the $n$-dimensional volume of integration. Generally, the performance of the Monte Carlo estimate, \ie its convergence with increasing $N$, is best for functions whose absolute value does not vary appreciably over $V$. If this is not the case, the performance can be improved by importance sampling \cite{NumericalRecipes}, \ie drawing the jump time sample $\mathfrak t_i^n$ from a suitable probability density $P(\mathfrak t^n)$ and weighing each realization $f(\mathfrak t^n_i)$ correspondingly,
\begin{equation}
 \int_V f(\mathfrak t^n) \mathd \mathfrak t^n \approx \frac 1 N \sum_{i=1}^N \frac {f(\mathfrak t^n_i)}{P(\mathfrak t^n_i)}.
\end{equation}
The performance of the Monte Carlo estimate is ideal for $P(\mathfrak t^n) \propto |f(\mathfrak t^n)|$. Translating the algorithm into the language of density matrices, we substitute $f$ with the integrand $\rho_t^{(\mathfrak t^n)}$ in Eq.~\eqref{mcintterm} and the modulus $|f|$ with the trace $\tmop{Tr}[\rho_t^{(\mathfrak t^n)}$].

Compare this to the standard quantum trajectory method \cite{Breuer2002,Knight1998}, which is similarly based on the norm of the state: There one generally propagates the initial (pure) state with $\msf H^\text{eff}$ from time 0 to $t$ in small steps $\Delta t$, such that the probability for a jump is based on the norm decay of the state at that instant. This implies that one needs to calculate a (time-local) jump probability after each step. In the present Monte Carlo integration method, in contrast, the $\mathfrak t^n_i$ are drawn from a preexisting (global) jump time distribution $P(\mathfrak t^n)$ and $\rho_0$ is propagated with the thereby determined series of transformations in one step.

For simplicity, we here choose to draw all jump times $t_1,\ldots, t_n$ independently, and from a single probability distribution $p(t)$. Subsequently, we sort them in the required ascending order, $0 < t_1 < \ldots t_n < t$, such that we have $n!$ possibilities to draw the same sequence $\mathfrak t^n$ of jump times. The joint probability density $P(\mathfrak t^n)$ is therefore given by
\begin{equation}
 P(\mathfrak t^n) = n! \prod_{j=1}^n p(t_j).
 \label{totjumptdist}
\end{equation}

This simple choice for $P(\mathfrak t^n)$ is already sufficient to study the convergence properties of the jump expansion for the open quantum problems discussed below. In particular, it leads to an ideal performance for the harmonic oscillator at $T=0$, for collisional decoherence, and for the continuous measurement in case of $\bs \alpha = 0$ (see Sec.~\ref{sec:casestudies}). However, Eq.~\eqref{totjumptdist} does not guarantee ideal performance in general. Since calculating the ideal global jump time distribution $P(\mathfrak t^n) \propto \tmop{Tr}[\rho_t^{(\mathfrak t^n)}]$ explicitly  may be numerically expensive, one idea to improve the performance of the Monte Carlo integration method for general master equations is as follows. One can approximate the ideal $P (\mathfrak t^n)$ successively by making an initial guess such as $P (\mathfrak t^n) \approx \text{const.}$ and then update $P (\mathfrak t^n)$ with the value of $\tmop{Tr} [\rho_t^{(\mathfrak t_i^n)}]$ obtained in each iteration of the integration method. Another interesting question, which is beyond the scope of the present article, is how the performance of the above Monte Carlo estimate of $\rho_t$ compares to that of quantum trajectory methods. While for quantum trajectory methods the (possibly complicated) calculation of $P (\mathfrak t^n)$ is not an issue, the possibility to propagate the initial state in a single step may well favor the classical Monte Carlo integration method.

\section{Case Studies}
\label{sec:casestudies}

As testbeds for the described numerical scheme and to asses the convergence of the adaptive jump expansion, we consider three exemplary Markovian open quantum problems in the following: (i) the damped harmonic oscillator, (ii) spatial decoherence of a particle, and (iii) a continuous quantum measurement with feedback. All three have been studied intensively over the past decades \cite{Breuer2002,Joos2003, Gisin1995, Strunz2002, Caldeira1983, Joos1985, Hornberger2009c, *Hornberger2010, Diosi2000, Hornberger2003a, Hornberger2003, Hornberger2009, Haroche2007,Fleming1990}, and were found to display many different facets of Markovian quantum dynamics such as decoherence, dissipation, thermalization, and pointer states. Moreover, these examples involve both discrete and continuous Hilbert spaces and feature finite and continuous sets of Hermitian or non-Hermitian jump operators. Benchmarking the optimal resummation in these case studies can hence be taken as an indication of its wide applicability.

We will see that the optimal resummation generally converges within the lowest two to five orders. A comparison with the convergence properties without resummation (\ie for $\bs \alpha = 0$) conveys an idea of the \emph{performance gain} achieved by the resummation. The jump expansion with $\bs \alpha = 0$ converges only within the lowest 20 to 80 orders depending on the considered model system, with one notable exception discussed below. These results were obtained for specific, physically motivated parameters.

To compare the performance of the optimal resummation to the jump expansion without resummation, we calculate the fidelity between the numerically exact solution of the master equation  $\rho_\tau$ and its expansion up to $k$-th order. We will choose the time $\tau$ of comparison to be around the intrinsic incoherent time scale of the master equation. The jump expansion is calculated using the classical Monte Carlo integration described in Sec.~\ref{sec:numericalimp}, while the numerically exact $\rho_\tau$ is obtained by numerically propagating the respective master equation in a finite basis. For the classical Monte Carlo method we use the jump time distribution $p(t) \propto \sum_j \tmop{Tr}[\msf L_j^\dag \msf L_j \rho_t]$ if $\bs \alpha = 0$ and a uniform distribution for the optimal $\tilde{\bs \alpha}$, see Eq.~\eqref{totjumptdist}.

The fidelity
\begin{equation}
 \mathcal F(\sigma,\rho) = \tmop{Tr}\sqrt{\sqrt \sigma \rho \sqrt \sigma}
\end{equation}
quantifies the distinguishability of two density matrices $\sigma$ and $\rho$. It ranges between zero and unity and is maximal for $\sigma = \rho$. The fidelity between $\rho_\tau$ and the expansion up to $k$-th order reads
\begin{equation}
 \mathcal F_k \equiv \mathcal F\left(\rho_\tau,\mathcal N_k^{-1} \sum_{n=0}^k \rho_\tau^{(n)}\right),\label{kfid}
\end{equation}
with $\mathcal N_k = \tmop{Tr}\sum_{n=0}^k \rho_\tau^{(n)}$. For a highly convergent expansion we expect that already the lowest order terms are very close to the true $\rho_\tau$, such that $\mathcal F_k$ rapidly reaches values close to unity. In contrast, the lowest orders of an expansion with poor convergence differ appreciably from the true $\rho_\tau$, such that $\mathcal F_k$ should increase more slowly with $k$. $\mathcal F_k$ is therefore a good measure of the convergence of the considered expansion.

\subsection{The Damped Harmonic Oscilllator}
\label{sec:dho}

\begin{figure*}[t]
   \centering
   \includegraphics[width=1\textwidth]{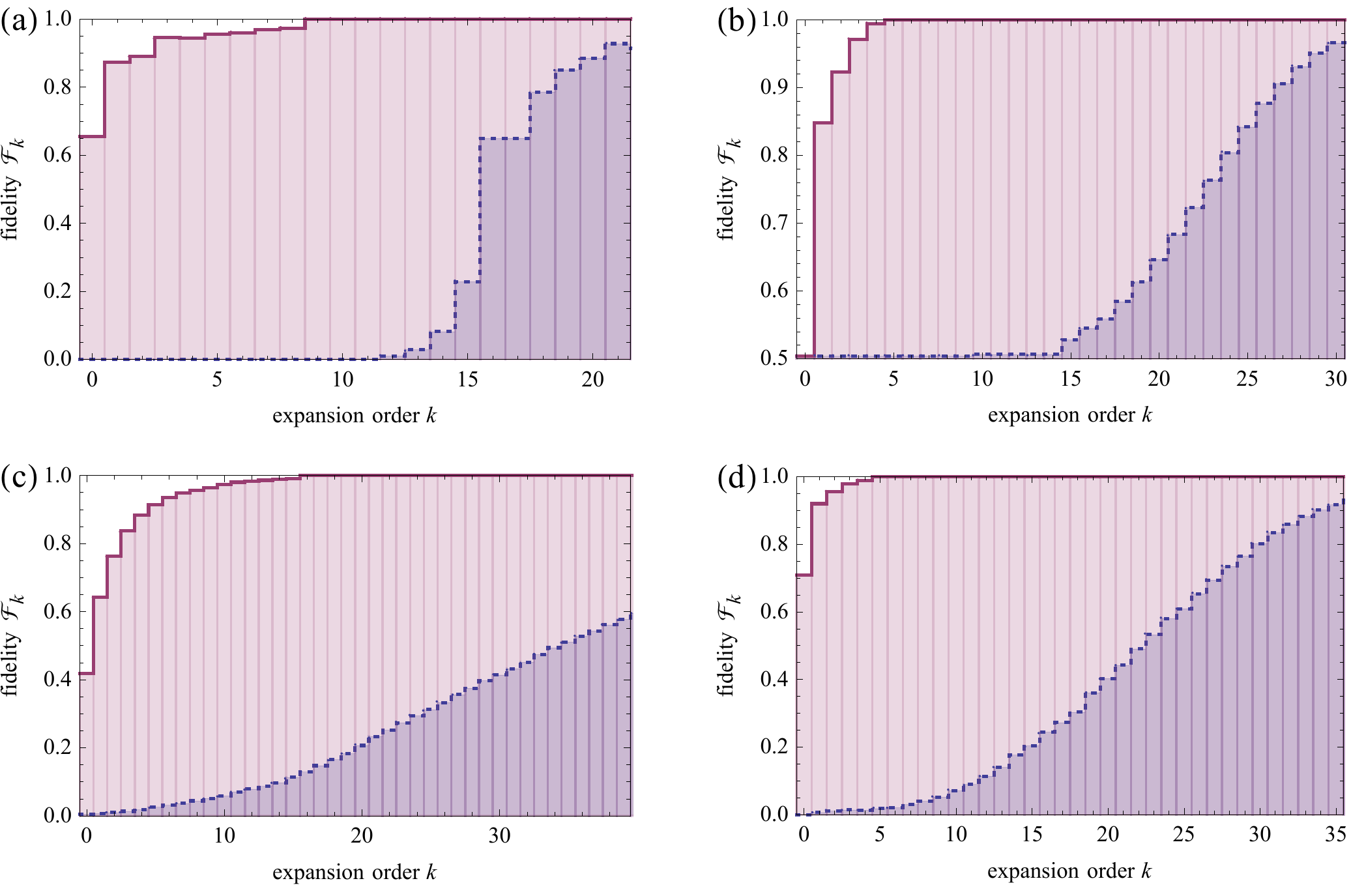}
   \caption{(Color online) Convergence of the jump expansion of $\rho_\tau$ without resummation, $\bs \alpha = 0$, (dashed blue line) and for the optimal resummation $\tilde{ \bs\alpha}$, Eq.~\eqref{optalpha} (solid purple line). The convergence is quantified by the fidelity $\mathcal F_k$ between $\rho_\tau$ with the expansion $\rho_\tau^{(0)} + \ldots + \rho_\tau^{(k)}$ up to $k$-th order, see Eq.~\eqref{kfid}. The four panels represent different master equations: (a) the damped harmonic oscillator, Eq.~\eqref{qomeq}, for a finite temperature corresponding to a thermal occupation $N_\text{th}=0.5$, and (b) for zero temperature, \ie $N_\text{th}=0$, (c) the diffusive limit of quantum Brownian motion, Eq.~\eqref{lincoup}, and (d) the non-selective measurement with feedback, Eq.~\eqref{nonselectmeas}. One observes that in all cases the resummation leads to a highly improved convergence.}
   \label{fig:fidelities}
\end{figure*}

We first consider a quantum harmonic oscillator with frequency $\omega$ and Hamiltonian $\msf H = \hbar \omega \msf a^\dag \msf a$, weakly coupled to a thermal bath of harmonic oscillators $\msf H_\text{b} = \int_0^\infty \hbar \omega' g(\omega') \msf b^\dag(\omega') \msf b(\omega') \mathd \omega'$. Its coarse-grained dynamics is described by a Markovian master equation with two jump operators \cite{Breuer2002},
\begin{align}
 \partial_t \rho_t &= -i \omega \left[ \msf a^\dag \msf a,\rho_t \right] -\gamma (N_\text{th}+1) (\msf a \rho_t \msf a^\dag - \{\msf a^\dag \msf a, \rho_t\}) \nonumber\\
& \phantom{=} - \gamma N_\text{th}(\msf a^\dag \rho_t \msf a - \{\msf a \msf a^\dag, \rho_t\}).
 \label{qomeq}
\end{align}
Here, the damping constant $\gamma\equiv 2 \pi g(\omega)|\kappa(\omega)|^2$ is determined by the density of modes $g$ of the bath and its coupling $\kappa$ at the frequency of the harmonic oscillator. $N_\text{th}$ stands for the thermal occupation number at frequency $\omega$, $N_\text{th} = \tmop{Tr} [\msf b^\dag(\omega) \msf b(\omega)  e^{-\msf H_\text{b}/k_B T}]/\tmop{Tr} [e^{-\msf H_\text{b}/k_B T)}]$. Equation~\eqref{qomeq} has a unique fixed point, given by the thermal state $\rho_T = \exp(-\msf H/k_B T)$, which is reached on the dissipation time scale $\tau_\text{dis} = 1/\gamma$.

In Fig.~\ref{fig:fidelities} (a) we plot the convergence of the jump expansion for the damped harmonic oscillator at a finite temperature, $N_\text{th}=0.5$. We choose $\omega / \gamma = 2$ and $\tau = 6 \tau_\text{dis}$, and use the exemplary initial state $\ket{\psi_0} = (\ket{19} + \ket{18} + \ket{17} + \ket{16})/2$. The convergence is quantified by the fidelity $\mathcal F_k$ given by Eq.~\eqref{kfid}. As expected, $\mathcal F_k$ increases rapidly for the optimal resummation, reaching values around 0.95 after $k = 3$ orders (solid purple line). This behavior indicates a rapid convergence of the jump expansion, as discussed above. The jump expansion without resummation, in contrast, needs $k\approx 19$ orders to attain comparable values of $\mathcal F_k$ (dashed blue line), indicating a much slower convergence.

\subsubsection*{Zero Temperature Limit}

A particularly important limiting case of Eq.~\eqref{qomeq} is a bath at zero temperature. The mean occupation number $N_\text{th}$ vanishes for $T=0$, which eliminates the jump operator $\msf a^\dag$, and the absolute ground state $\ket 0 \bra 0$ becomes the unique fixed point or steady state of Eq.~\eqref{qomeq}. The dynamics under the remaining jump operator $\msf a$ is now analytically solvable in terms of the eigenstates of $\msf a$, \ie the coherent states $\ket \beta$.

An initial coherent state $\rho_0 = \ket{\beta(0)}\bra{\beta(0)}$ remains a pure coherent state $\rho_t = \ket{\beta(t)}\bra{\beta(t)}$, spiraling towards the origin of the complex plane,
\begin{equation}
 \beta(t) = \beta(0) \exp[(- \gamma/2 + i\omega)t].
\end{equation}
Its mean energy $\bra{\beta(t)}\msf H\ket{\beta(t)} = |\beta(0)|^2 \exp(-\gamma t)$ decays exponentially on the dissipation time scale $\tau_\text{dis} = 1 / \gamma$  until the steady state is reached.

An initial superposition of two coherent states $\rho_0 = \ket{\psi_0}\bra{\psi_0}$ with $\ket{\psi_0} = c_1\ket{\beta_1(0)} + c_2 \ket{\beta_2(0)}$ decays into a probabilistic mixture of the coherent states $\ket{\beta_1(t)}$ and $\ket{\beta_2(t)}$, with probabilities given by the absolute squares of the initial expansion coefficients. The full evolution reads
\setlength\multlinegap{0pt}
\begin{equation}
\begin{split}
\rho_t &= |c_1|^2\ket{\beta_1(t)}\bra{\beta_1(t)} + |c_2|^2\ket{\beta_2(t)}\bra{\beta_2(t)} \\
  &\phantom{=} + c_1 c_2^\ast D(t)\ket{\beta_1(t)}\bra{\beta_2(t)} + c_1^\ast c_2 D(t)\ket{\beta_2(t)}\bra{\beta_1(t)},
  \end{split}
 \label{psmixture}
\end{equation}
where the decay of the coherences is governed by $D(t) = \exp[z (1 - e^{-\gamma t})]$ with the complex coefficient $z = -|\beta_1(0) - \beta_2(0)|^2/2 + i \tmop{Im}[\beta_1(0)\beta_2(0)^\ast]$. For small times $t \ll \gamma^{-1}$ one therefore obtains an exponential decay on the decoherence time scale $\tau_\text{dec} = 2/(\gamma|\beta_1(0) - \beta_2(0)|^2)$, which is much shorter than $\tau_\text{dis}$ if $\beta_1(0)$ and $\beta_2(0)$ are distinguishable.

After about the decoherence time $\tau_\text{dec}$ we expect that each record conditioned branch of the optimally convergent jump expansion is close to an eigenstate of the jump operator $\msf a$. In Sec.~\ref{sec:optimalresummation} we saw that the jump rates vanish for eigenstates of the jump operators, which indicates that the optimized jump expansion converges on the time scale $\tau_\text{dec}$. The jump expansion without resummation, in contrast, is expected to converge on the much longer time scale $\tau_\text{dis}$ on which the steady state is reached, therefore admitting a much higher number of jumps.

Indeed, in Fig.~\ref{fig:fidelities} (b) we see that the difference of convergence between the optimal resummation and the jump expansion without resummation is even more pronounced than for the finite temperature case (Fig.~\ref{fig:fidelities} (a)): while the former converges within $k\approx 3$ orders (solid purple line), the latter needs $k\approx 30$ order to reach comparable values of $\mathcal F_k$ (dashed blue line). We use $\omega / \gamma = 2$ as before, and $\tau = 2 \tau_\text{dis}$. Initially, the system is in a balanced superposition of two coherent states $\beta_1(0) = 0$ and $\beta_2(0) = 6$. A peculiar feature of this initial state is that $\mathcal F_k$ is never below $0.5$ both for the optimal $\tilde{\bs \alpha}$ and for $\bs \alpha = 0$.

\begin{figure*}[t]
   \centering
   \includegraphics[width=1\textwidth]{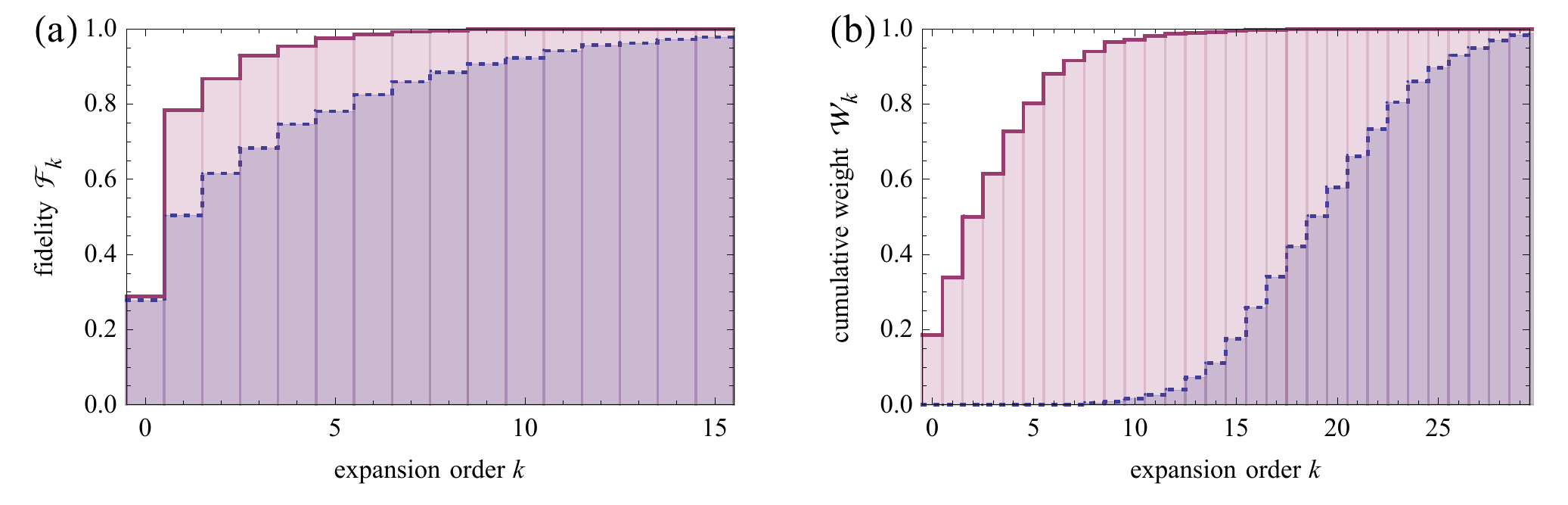}
   \caption{(Color online) Convergence of the jump expansion for collisional decoherence, Eq.~\eqref{coldeco}, for the optimal resummation $\tilde{\bs \alpha}$ (purple solid line) and without resummation, $\bs \alpha =0$ (blue dashed line). In (a) we plot the fidelity $\mathcal F_k$ analogous to Fig.~\ref{fig:fidelities}, whereas (b) shows the cumulative sum of the weights of the expansion terms up to $k$-th order, $\mathcal W_k  = w_0(t) + \ldots + w_k(t)$. The comparison of (a) and (b) shows that a rapid increase of the fidelity even for $\bs \alpha =0$ does not imply a strong convergence of the jump expansion. It is rather a special feature of collisional decoherence (compare to $\mathcal F_k$ for different exemplary master equations, Fig.~\ref{fig:fidelities}).}
   \label{fig:coldeco}
\end{figure*}

\subsection{Spatial Decoherence}

\subsubsection*{Pure Diffusion}

The quantum Brownian motion of a freely moving particle can be described by coupling it linearly to a thermal bath of harmonic oscillators (see previous example)
\cite{Caldeira1983}.
The diffusive and high-temperature limit of this so-called Caldeira-Legget model is governed by the master equation \cite{Joos1985}:
\begin{equation}
 \partial_t \rho_t = -\frac i \hbar \left[\frac {\msf p^2}{2m} ,\rho_t \right] - \frac{4\pi \gamma}{\Lambda_\text{th}^2} [\msf x,[\msf x, \rho_t]].
 \label{lincoup}
\end{equation}
Here $\Lambda_\text{th} = 2 \pi \hbar^2 / m k_B T$ is the thermal de Broglie wavelength of the Brownian particle. Equation~\eqref{lincoup} describes spatial decoherence of the particle on a timescale proportional to $1/\gamma$, but does not include frictional dissipation. Due to the absence of dissipation, Eq.~\eqref{lincoup} has no fixed point.

Figure~\ref{fig:fidelities} (c) confirms the strong convergence gain of the optimal resummation, already observed in the damped harmonic oscillator, for spatial decoherence. Again, the optimal resummation converges after a few orders, here around $k \approx 5$, whereas the jump expansion without resummation requires around $k \approx 80$ orders to attain comparable values of $\mathcal F_k$ (not shown on the scale of the plot). Here we used $\hbar/8\pi m \gamma = 15 \Lambda_\text{th}^2$ in Eq.~\eqref{lincoup} and set $\gamma \tau = 1/3$. The initial state is taken to be a Gaussian wave packet centered at $x_0 = 0$ with width $\sigma_x = 3 \Lambda_\text{th}$ and mean wave number $k_0 = 2 / \Lambda_\text{th}$.

\subsubsection*{Collisional Decoherence}

A different approach to the decoherence of a Brownian particle, called collisional decoherence \cite{Joos1985, Fleming1990, Hornberger2003a, Hornberger2009}, has been confirmed in molecular matter wave interference experiments \cite{Hornberger2003, Hornberger2004}. It describes the interaction of the particle with its environment as a number of separate, uncorrelated collision events. In the limit of a large mass ratio, elastic collisions lead to momentum kicks $\msf L_q = \exp[i q \msf x/ \hbar]$ that leave the particle position unaffected \cite{Hornberger2009b}. The master equation, giving again rise to dissipationless dynamics, reads
\begin{equation}
 \partial_t \rho_t = -\frac i \hbar \left[\frac {\msf p^2}{2 m},\rho\right] + \gamma \int G(q) \left(e^{i q \msf x/\hbar} \rho_t e^{-i q \msf x/\hbar} - \rho_t \right)\mathd q.
 \label{coldeco}
\end{equation}
Here, $\gamma$ is the collision rate of particle and background gas and $G(q)$ is the normalized momentum transfer distribution. Both can be related to microscopic quantities of the environment gas \cite{Fleming1990, Hornberger2003a}; we take $G(q)$ to be a Gaussian of width $\sigma_G$, centered around $q = 0$.

In Fig.~\ref{fig:coldeco} (a) we plot the fidelity $\mathcal F_k$ at time $\gamma \tau = 20$ for $2m \hbar \gamma = 4 \sigma_G^2$ and an initial Gaussian wave packet centered at $x_0 = 0$ with $\sigma_x = 3 \hbar / \sigma_G$. As in the previous examples, the optimal resummation converges rapidly within the lowest $k \approx 4$ orders (solid purple line). The difference here is that also for $\bs \alpha = 0$ the fidelity shows a rapid increase at small $k$ (dashed blue line). While one still needs around $k \approx 13$ orders to reach a fidelity of 0.95, we observe that already the lowest orders yield a fairly good estimate for $\rho_\tau$. Therefore, we also plot the cumulative weight distribution $\mathcal W_k  = w_0(\tau) + \ldots + w_k(\tau)$ in Fig.~\ref{fig:coldeco} (b). Note here that the weights $w_k(\tau)$ are not only the actual subject of optimization by the resummation, they are also the only available criterium to estimate whether the jump expansion of an unknown state $\rho_\tau$ has converged. Figure~\ref{fig:coldeco} (b) shows that the lowest order weights are strongly suppressed for $\bs \alpha = 0$ and that the convergence for the optimal $\tilde{\bs \alpha}$ is clearly superior. The unexpectedly rapid increase of $\mathcal F_k$ with $k$ despite a rather slow convergence of the expansion seems to be a special feature of collisional decoherence, Eq.~\eqref{coldeco}. It was not observed in the other exemplary master equations, see Fig.~\ref{fig:fidelities}. It underscores that a rapid increase of fidelity does not imply a strong convergence of the jump expansion.

\subsection{Continuous, Nonselective Measurements}

A general quantum measurement can be described in terms of a set of measurement operators $\msf M_{j,k}$ and effects $\msf F_j = \sum_k \msf M_{j,k}^\dag \msf M_{j,k}$ with $\sum_j \msf F_j = \mathbbm{1}$ \cite{Breuer2002}. This allows one to extend the formal projective measurement setting to realistic measurements with finite resolution and classical uncertainty and to incorporate a feedback conditioned on the measurement outcomes. A general measurement of a system in state $\rho$ yields the outcome $j$ with probability $p_j = \tmop{Tr}[\msf F_j \rho]$ and it transforms the state to
\begin{equation}
 \rho_j = \frac{\sum_k \msf M_{j,k} \rho \msf M_{j,k}^\dag}{\tmop{Tr}[\msf F_j \rho]}.
 \label{genmeas}
\end{equation}
If the measurement is performed probabilistically with rate $\gamma$ and the outcomes are disregarded, we talk about a continuous, nonselective measurement. The dynamics of the measured system is then governed by the master equation
\begin{equation}
 \partial_t \rho_t = -\frac i \hbar [\msf H, \rho_t] + \gamma \left(\sum_{j,k} \msf M_{j,k} \rho_t \msf M_{j,k}^\dag - \rho_t\right).
 \label{nonselectmeas}
\end{equation}

\subsubsection*{Measurement Setup in Cavity QED Experiments}

For the sake of concreteness, let us consider a particularly well studied setup allowing one to implement an unsharp quantum measurement with feedback \cite{Haroche2007, Haroche2011}. The measured system is an electromagnetic field mode in a cavity ($\msf H = \hbar \omega \msf a^\dag \msf a$) which is initially in the unknown pure state $\ket{\psi_0} = \sum_n c_n\ket n$. The cavity is probed indirectly by sending individual two-level atoms prepared in $\ket + \propto \ket 0 + \ket 1$ through the cavity. Through their interaction with the cavity, the atoms acquire an $n$-dependent phase $\ket{\varphi} = \sum_n c_n \ket{n} \ket{+_n}$, with $\ket{\pm_n} = \exp[\pi i (n/2d) \msf \sigma_z]\ket \pm$. After the atoms have left the cavity, one can subject them to a projective measurement in either of the $k$-labelled measurement bases $\ket{\pm_k}$, with $k = 1, \ldots, d$. A projection onto a specific basis $k$ then implements the measurement operators $\msf M_{\pm,k} = \sum_{n=0}^d \bra{\pm_k} +_n\rangle\ket n \bra n$ for the cavity. While for a given $k$, $\msf M_{+,k}$ and $\msf M_{-,k}$ define a valid measurement of the cavity state, one can switch between different $k$ in consecutive cavity measurements, \eg to increase the amount of extracted information. For the present continuous nonselective measurement, we choose the exemplary parameter $d = 19$ and assume that measurements with $k=0$ and $k=10$ occur with equal probability. We therefore have the four measurement operators $\msf M_{+,0}$, $\msf M_{-,0}$, $\msf M_{+,10}$, and $\msf M_{-,10}$ in Eq.~\eqref{nonselectmeas}.

Besides a mere measurement of the photon number, this setup allows one to apply a feedback conditioned on the measurement outcomes and thereby stabilize a specific number state \cite{Haroche2011}. Here we use the addition of a photon $\msf a^\dag$ as feedback operation\footnote{In the cited experiment, the feedback operation was the displacement $\msf D(\beta)$ by an optimal, state-dependent displacement vector $\beta$. We apply $\msf a^\dag$ since the corresponding master equation is formally clearer. An experimental realization of the feedback operation $\msf a^\dag$ could be based, for example, on a second, in-resonance two level atom which deposits a quantum with a high probability.} and apply it only when we register the outcome $+,0$. We thus have $\msf M_{+,0} = \sum_{n=0}^d \bra{+_0} +_n\rangle\ket{n + 1} \bra n$, while all other measurement operators remain as specified. This feedback stabilizes the cavity in state $\ket{19}$, since the probability $p_{+,0} = \tmop{Tr} [\msf M_{+,0}^\dag \msf M_{+,0} \rho]$ vanishes for $\rho = \ket{19}\bra{19}$ and hence no further feedback operation is applied.

In Fig.~\ref{fig:fidelities} (d) we plot the convergence of the jump expansion for the nonselective measurement \eqref{nonselectmeas} with the measurement operators defined above. Here, we chose the measurement rate $\gamma = 2 \omega$, the initial coherent cavity state $\ket{\psi_0} = e^{-1/2}e^{2 \msf a^\dag} \ket 0$, and the time $\gamma \tau = 40$. Again, the high convergence of the optimal resummation observed in the previous examples is confirmed. It converges within $k \approx 2$ orders, whereas one needs to take into account $k\approx 40$ orders for $\bs \alpha = 0$ to obtain a faithful estimate of $\rho_\tau$.

\section{Conclusions}
\label{sec:conclusion}

We have seen in the first part of this article that one can obtain a formal expansion of Markovian quantum dynamics into periods of continuous evolution and discontinuous, random jumps by decomposing the generator into a sum of two parts. Different possible expansions are parametrized by tuples of complex numbers $\bs \alpha$, reflecting the invariance of the master equation under a transformation of its jump operators. Casting the weights of the jump terms into an $\bs \alpha$-dependent form, the convergence of the series can be optimized for a suitable choice $\tilde{\bs \alpha}$. Since the latter is conditioned on the complete information about all previous jumps, its implementation amounts to an adaptive, optimally convergent resummation of the jump expansion.

The dependence of $\tilde{\bs \alpha}$ on the complete jump record suggested that one can obtain suboptimal resummations that are structurally simpler by reducing the information contained in $\tilde{\bs \alpha}$. This may facilitate the analytical treatment of open quantum dynamics \cite{Lucas2013}. Along these lines one might also optimize the jump expansion for rather different purposes, such as incoherent control tasks~\cite{Wiseman2005}.

Moreover, we showed that the jump rates (or entropy production rates) are minimized for the optimal resummation. This minimality is a defining property of the so-called pointer states \cite{Zurek2003}, \ie the basis states distinguished by the open quantum dynamics.
In accord with empirical observations, we showed that the jump rates in the optimal resummation are particularly low or can even vanish altogether if the jump operators define a preferred (pointer) basis. In view of the fact that the pointer states of most realistic open systems exhibit small but non-vanishing jump rates, the mixed state terms of the resummed jump expansion may reveal new aspects about the quantum-to-classical transition not captured by current methods based on pure state quantum trajectories \cite{Diosi2000,Hornberger2009c, *Hornberger2010}.

In the second part, the optimally resummed jump expansion was implemented numerically for a number of well-known master equations. We discussed how to compute the jump terms straightforwardly by means of classical Monte Carlo integration with importance sampling, and we found convergence within the lowest orders, irrespective of the formal differences of the studied master equations (single vs.\ continuous set of jump operators, Hermitian vs.\ non-Hermitian jump operators, finite vs.\ infinite dimensional Hilbert space).

The fact that high convergence was observed in all cases underscores that the resummation method has the potential to yield analytic approximations for a wide class of open quantum problems, extending beyond the examples discussed in Ref.~\cite{Lucas2013}. It also implies a viable method for the efficient numerical simulation of master equations, which works by maximally biasing the weights of the constituent terms and then computing them successively, starting from the most important one. It thus complements the standard quantum trajectory approach, which approximates the overall mixed state by a single indiscriminate sum of pure states.

Since the method advances our abilities on both the analytical and the numerical side, it may well become a versatile tool for the study of Markovian open systems. Offering a new perspective on the fundamental difference between quantum and classical dynamics, it may lead to analytical models of open dynamics which are relevant for tasks such as incoherent quantum control.

\end{document}